\documentclass[11pt]{article}

\usepackage{graphicx} 
\usepackage[numbers,super]{natbib}
\usepackage{bm}
\usepackage{amsmath}
\usepackage[a4paper, top=1in, bottom=1in, left=1in, right=1in]{geometry}
\usepackage{setspace}
\usepackage{rotating}
\usepackage{booktabs}
\usepackage{array}
\usepackage{hyperref}
\usepackage{xcolor}
\usepackage{amsfonts}

\hypersetup{
    colorlinks=true,
    linkcolor=blue,
    citecolor=blue,
    filecolor=magenta,      
    urlcolor=blue,
    pdfpagemode=FullScreen,
    }

\newcommand{\bbeta}{{\bm \beta}}
\newcommand{\bgamma}{{\bm \gamma}}
\newcommand{\biw}{\boldsymbol{w}}
\newcommand{\bix}{\boldsymbol{x}}

\title{Spatially Varying Coefficient Models for Estimating Heterogeneous Mixture Effects}
\author{
Jacob Englert$^1$ and Howard Chang, PhD$^{1,2}$ \\
$^1$Department of Biostatistics and Bioinformatics, Emory University \\
$^2$Gangarosa Department of Environmental Health, Emory University}
\date{February 2025}

\begin{document}

\maketitle

\begin{abstract}
    Recent studies of associations between environmental exposures and health outcomes have shifted toward estimating the effect of simultaneous exposure to multiple chemicals. Summary index methods, such as the weighted quantile sum and quantile g-computation, are now commonly used to analyze environmental exposure mixtures in a broad range of applications. These methods provide a simple and interpretable framework for quantifying mixture effects. However, when data arise from a large geographical study region, it may be unreasonable to expect a common mixture effect. In this work, we explore the use of a recently developed spatially varying coefficient model based on Bayesian additive regression trees to estimate spatially heterogeneous mixture effects using quantile g-computation. We conducted simulation studies to evaluate the method's performance. We then applied this model to an analysis of multiple ambient air pollutants and birthweight in Georgia, USA from 2005-2016. We find evidence of county-level spatially varying mixture associations, where for 17 of 159 counties in Georgia, elevated concentrations of a mixture of PM$_{2.5}$, nitrogen dioxide, sulfur dioxide, ozone, and carbon monoxide were associated with a reduction in birthweight by as much as -16.65 grams (95\% credible interval: -33.93, -0.40) per decile increase in all five air pollutants.
\end{abstract}

In 2022, the National Center for Health Statistics reported that an estimated 8.60\% of infants born in the United States had low birthweight (less than 2,500 grams).\citep{osterman_births_2024} Low birthweight has a strong association with infant mortality and morbidity. In the same year, infant mortality rate in the United States was 42.36 per 1,000 live births among low birthweight infants, compared to just 2.10 per 1,000 live births among infants greater than 2,500 grams.\citep{ely_infant_2024} Identifying risk factors of reduced birthweight, particularly those due to modifiable environmental risk factors, is an important research priority.

In environmental and perinatal epidemiology, there is a rich literature supporting the associations between various air pollutants and birth outcomes including, but not limited to, reduced birthweight. \citep{lamichhane_meta-analysis_2015, li_association_2017, stieb_ambient_2012, sun_associations_2016} These studies have commonly identified associations between low birthweight and elevated concentrations of PM$_{2.5}$, nitrogen dioxide, sulfur dioxide, ozone, and carbon monoxide, among others. These associations have previously been reported in Atlanta, Georgia.\citep{darrow_ambient_2011, strickland_associations_2019}

In the past, studies of association between air pollution and health outcomes have utilized single-exposure models. While useful, these models may be inadequate for describing the combined effect of multiple air pollutants that individuals are simultaneously exposed to. In recent years, research has shifted toward developing and applying mixture models that attempt to quantify this joint association between multiple exposures and health. Of the many modeling strategies introduced, quantile g-computation (QGCOMP) proposed by \citet{keil_quantile-based_2020} has been the most widely used approach in population-based epidemiologic studies. QGCOMP is favored for its simple definition, computational speed and interpretation of the overall mixture effect, as well as its straightforward implementation via the well-maintained \texttt{qgcomp} R package.\citep{keil_qgcomp_2019}

In many studies of environmental mixtures where QGCOMP or alternative approaches might be used, it is common to have health and exposure data acquired from a large geographical study region. While compiling data from all regions within the study area increases sample size and the ability to detect small mixture effects, it also provides an opportunity to explore spatial heterogeneity in health effects.

In this work, we consider a varying coefficient model based on Bayesian additive regression trees (BART) \citep{chipman_bart_2010, deshpande_vcbart_2024} to estimate spatially heterogeneous mixtures effects within the QGCOMP framework. BART is a flexible modeling approach that has consistently performed well on a variety of prediction, classification, and causal inference tasks.\citep{chipman_bart_2010, hill_bayesian_2020, hahn_bayesian_2020} An additional benefit of using BART is that, unlike most other machine learning models, BART is fully Bayesian and thus offers natural uncertainty quantification via the posterior distribution. We conduct a simulation study to evaluate the method in the presence of spatially varying mixture effects, including a comparison with spatially varying coefficient models from disease mapping, and then apply the method to an analysis of birthweight from vital records in the state of Georgia.

\section*{Data}

\subsection*{Air Pollution Data}
We considered five air pollutants: fine particulate matter with diameter 2.5 µm and smaller (PM$_{2.5}$, 24-hr average, µm/m$^3$), nitrogen dioxide (NO$_2$, 1-hr max, ppb), sulfur dioxide (SO$_2$, 1-hr max, ppb), ozone (O$_3$, 8-hr max, ppm), and carbon monoxide (CO, 1-hr max, ppb). Daily estimates of the concentrations of each pollutant were derived from a data fusion model which utilized simulations from the Community Multiscale Air Quality Model and monitoring data from the Environmental Protection Agency's Air Quality System database.\citep{senthilkumar_using_2022} The original data product is available at a 12 km gridded spatial resolution. We used area-weighted averaging to obtain exposures at the ZIP code level.

\subsection*{Health Data}
We obtained birth records from the Office of Health Indicators for Planning, Georgia Department of Public Health. We restricted the data to only include singleton pregnancies with gestational age greater than 27 weeks and an estimated date of conception between January 1st, 2005 and December 31st, 2016. There were a total of 1,468,531 births meeting this criteria. Additional covariates collected on the birth mothers included age (years), race, level of educational attainment, marital status, and parity. Pregnancy-wide air pollution exposures were estimated by linking maternal residential address ZIP code and calculating the average concentration of each pollutant from the date of conception to the date of birth.

\section*{Methods}
\label{sec:methods}

\subsection*{Review of Quantile g-Computation for Mixture Modeling}

The goal of QGCOMP, like its predecessor weighted quantile sum (WQS), is to provide a more interpretable mixture effect. For this reason, QGCOMP is sometimes referred to as a \emph{summary index} method. This stands in contrast to \emph{response surface} methods, such as Bayesian kernel machine regression (BKMR), which provide a more flexible approach to modeling complex exposure-response surfaces, but generally are not as easily implemented or interpreted. 

In the QGCOMP framework, the target parameter(s) quantify the expected change in the outcome due to an increase of one quantile in all exposures of interest. Because the implementation of QGCOMP leverages model fitting procedures from standard regression models, it can be run efficiently and has been extended to a variety of models. This is particularly important for studies which make use of administrative datasets, for which computationally burdensome methods such as BKMR are impractical.

Assuming the data arise from several study regions, the linear, additive quantile g-computation model is given by \eqref{eq:og-qgcomp}:
\begin{equation}
    \label{eq:og-qgcomp}
    Y_{ij} = \beta_0 + \bix_{ij}^T \bbeta + \biw_{ij}^T \bgamma + \epsilon_{ij} \qquad \epsilon_{ij} \overset{i.i.d.}{\sim} N(0, \sigma^2),
\end{equation}
where $\bix_{ij}$ is a $P_x$-vector of quantized exposures for observation $j$ in region $i$. Here, the term quantized exposure refers to an originally continuous exposure whose values have been recoded to $0, 1, \ldots, Q-1$, where the new value represents which of the $Q$ quantiles the original observed value belonged to. The model may also include $\biw_{ij}$, a $P_w$-vector containing confounders for adjustment for observation $j$ of region $i$ (note these are not generally quantized). Thus, each element of $\bbeta = (\beta_1, \ldots, \beta_{P_x})^T$ represents the expected change in the outcome for a one quantile increase in the corresponding exposure, while the elements of $\bgamma$ retain typical interpretations for regression coefficients.

The reasoning behind this treatment of the exposures of interest is to define a \emph{mixture effect} as $\Psi = \sum_{p=1}^{P_x} \beta_p$. When the model is linear and additive in terms of the quantized exposures, $\Psi$ represents the expected change in the outcome for a one quantile increase in every exposure. In this scenario, $\Psi$ also coincides with the slope parameter from a simple linear marginal structural model (MSM) in which the sole predictor, denoted $S_q$, represents the quantized exposure mixture. Specifically, $S_q$ takes on values $0, \ldots, Q-1$, corresponding to when all quantized exposures are simultaneously set to $0, \ldots, Q-1$. When framing the problem in this manner, $\Psi$ may alternatively be estimated via g-computation with the joint exposure quantile $S_q$. In this work we focus on the linear and additive model \eqref{eq:og-qgcomp}, but in general estimation of $\Psi$ via the MSM approach can be used when higher order terms are desired. Under certain assumptions, $\Psi$ might be interpreted as a causal parameter.\citep{keil_quantile-based_2020}


\subsection*{Spatially Varying Quantile g-Computation with BART}
\label{sec:sv-qgcomp}

We propose allowing the individual exposure coefficients $\bbeta$ to vary across space, which in turn implies the mixture effect $\Psi$ also varies across space. The result is the following model:
\begin{equation}\label{eq:sv-qgcomp}
    Y_{ij} = \beta_0 (z_{ij}) + \sum_{p=1}^{P_x} \beta_p(z_{ij}) x_{ij,p} + \biw_{ij}^T \bgamma + \epsilon_{ij} \qquad \epsilon_{ij} \overset{i.i.d.}{\sim} N(0, \sigma^2),
\end{equation}
where the intercept and quantized exposure coefficients depend on the spatial location of observation $ij$, $z_{ij}$. For simplicity we write $z_{ij} = z_i$, since we have already defined $i$ as indexing location. The local mixture effect specific to location $z_i$ is then defined as $\Psi(z_i) = \sum_{p=1}^{P_x} \beta_p(z_i)$.

There are a few reasons allowing for a spatially varying air pollutant mixture effect is warranted. For example, the mixture of PM$_{2.5}$ components may be different spatially due to differences in local emission sources and meteorology. Spatially varying population characteristics that impact the relationships between personal exposure and ambient concentration may also result in effect heterogeneity. When the target estimand is the overall mixture effect, differences may also be attributable to specific exposure levels due to a nonlinear exposure-response relationship. For example, NO$_2$ may be a more important component in the mixture for regions near highways where levels are high. Additionally, if the true exposure-response surface contains any interactions, then the effects of individual exposures and the overall mixture effect is likely different in regions with different exposure concentrations. Allowing for spatially varying weights allows for capturing locally linear mixture effects, even when the overall mixture effect is more complex.

\begin{figure}
    \centering
    \includegraphics[width=5.5in]{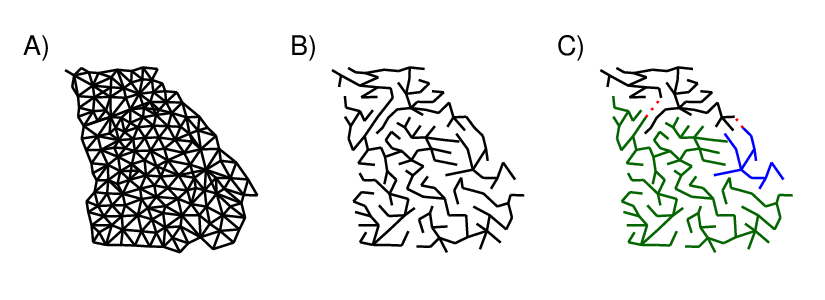}
    \caption{An illustration of VCBART's spatial branching process, applied to counties in Georgia. In panel A), edges are drawn between centroids of adjacent counties. In panel B), a random spanning tree is drawn from the graph in panel A). In panel C), three groups of contiguous counties are formed by randomly deleting two edges from the spanning tree in panel B).}
    \label{fig:spanning-tree-diagram}
\end{figure}

The spatially varying intercept and exposure coefficients in model \eqref{eq:sv-qgcomp} can be estimated in various ways; we suggest using BART priors for each of these parameters. \citet{deshpande_vcbart_2024} recently developed a varying coefficient BART (VCBART) model and demonstrated its use for studying time series of crime rates across census tracts in Philadelphia, Pennsylvania. When supplied with a list of which sub-regions are spatially adjacent to one another (i.e., share a border), VCBART uses efficient proposal mechanisms based on sampling spanning trees to repeatedly subdivide the study area into contiguous sub-regions which the data suggest are heterogeneous (see Figure \ref{fig:spanning-tree-diagram}).\citep{deshpande_flexbart_2024} This process is done separately for each of the spatially varying parameters in model \eqref{eq:sv-qgcomp}, which allows for different spatial clusters for each mixture component.

Estimation of the VCBART model is carried out using Markov chain Monte Carlo (MCMC), with each sample from the posterior distribution partitioning the study area differently. The posterior distribution of mixture effects might then be summarized for each location using their posterior means and 95\% credible intervals. BART priors function similarly to \emph{boosting}, as each tree contributes a small portion to the overall output, allowing for fine tuning of the spatial branching process. Consistent with other BART implementations, regularization priors are used to encourage homogeneity across the entire study area to prevent over-fitting.


\section*{Simulation Study}
\label{sec:sim-study}

\begin{figure}
    \centering
    \includegraphics[width=6.5in]{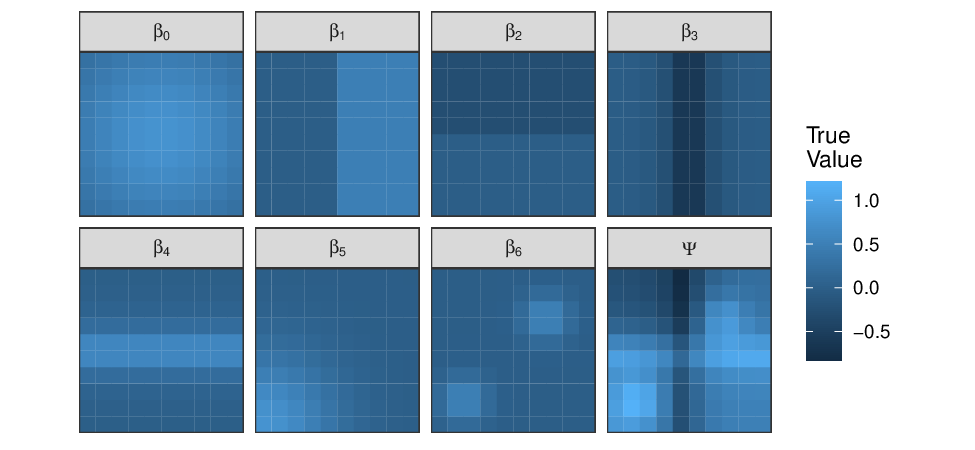}
    \caption{True spatially varying intercept ($\beta_0$), regression coefficients ($\beta_1$-$\beta_6$), and overall mixture effect ($\Psi$) surfaces used for the simulation study. All are defined on a 10 x 10 grid.}
    \label{fig:sim-true-surfaces}
\end{figure}

In this section we evaluate the ability of VCBART to estimate the spatially varying parameters of model \eqref{eq:sv-qgcomp}. We generate a spatially varying intercept, $\beta_0(z)$, and six spatially varying regression coefficients, $\beta_1(z), ..., \beta_6(z)$, across a 10 x 10 grid using various smooth and rigid functions (see the Supplementary Material for details of the functions). The surfaces are plotted in Figure \ref{fig:sim-true-surfaces}, along with the true mixture effect $\Psi(z)$.

For the simulation, the exposures are generated from a mean zero multivariate normal distribution with covariance $\rho \mathbf{J}_6 + (1-\rho) \mathbf{I}_6$. Each of the exposures is then quantized using $Q = 4$ quantile bins. Finally, the outcome is drawn from a normal distribution with some noise variance $\sigma^2$. Parameters varied during the simulation study include the sample size within each grid cell ($n \in \left\{10, 50, 100, 250\right\}$), the degree of correlation between exposures ($\rho \in \left\{0.0, 0.5, 0.8\right\}$), and the amount of noise ($\sigma \in \left\{0.1, 1\right\}$). Each parameter setting is run for $B = 200$ unique datasets.

We fit all VCBART models for the simulation using the \texttt{VCBART} R package publicly available on GitHub (\url{https://github.com/skdeshpande91/VCBART}). The default hyperparameter settings from the package are used, including 50 trees per BART ensemble. The most natural comparison might be the spatially varying coefficient (SVC) model, which uses Gaussian process (GP) priors in place of BART priors.\citep{gelfand_spatial_2003} Others have described the connection between BART and GP priors.\cite{linero_review_2017} However, due to the computational burden presented by GP priors, studies of areal data often make use of conditional autoregressive (CAR) priors.\citep{besag_spatial_1974} These reduce the dimension of the coefficients to the number of unique spatial locations and thus are more computationally convenient. For this reason, we compare VCBART to a model with proper Besag CAR priors on the intercept and each of the regression coefficients. We fit these models using the integrated nested Laplace approximation (INLA) available in the \texttt{INLA} R package (\url{https://www.r-inla.org}).

\begin{figure}
    \centering
    \includegraphics[width=5.5in]{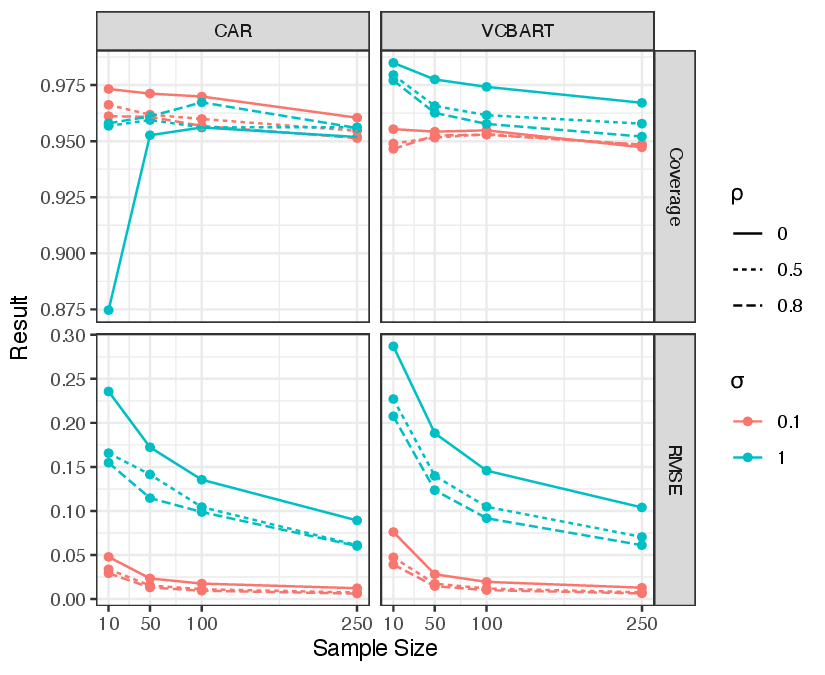}
    \caption{Global average 95\% credible interval coverage and root mean squared error (RMSE) for the mixture effect $\Psi$ using CAR and VCBART models in the simulation study. Global values are calculated using the following approach: $\text{Coverage} = \frac{1}{B} \sum_{b=1}^{B} \left(\frac{1}{100}\sum_{z} \mathbb{I}_{\left\{\Psi(z) \in [\hat{\Psi}(z)_{0.025}, \hat{\Psi}(z)_{0.975}] \right\}} \right)$, and $\text{RMSE} = \frac{1}{B} \sum_{b=1}^{B} \left(\frac{1}{100}\sum_{z} (\Psi(z) - \hat{\Psi}(z))^2 \right)$.}
    \label{fig:sim-stats-summmary-6}
\end{figure}

Figure \ref{fig:sim-stats-summmary-6} contains a summary of the simulation results for both the CAR and VCBART models in terms of 95\% credible interval coverage and root mean squared error (RMSE) for the 100 mixture effects (one for each grid cell). As the sample size increases, global coverage tends toward 95\% and RMSE decreases for both models. The CAR model has slightly better RMSE in small samples, but the difference is negligible in the $n = 250$ setting. In general, better global coverage and RMSE is observed when exposures exhibit stronger correlation. Despite this, the individual performance on any one of the spatially varying coefficients may decrease with increasing correlation (see Figures S1.1 and S1.2 in the Supplementary Material).

\begin{figure}
    \centering
    \includegraphics[width=6in]{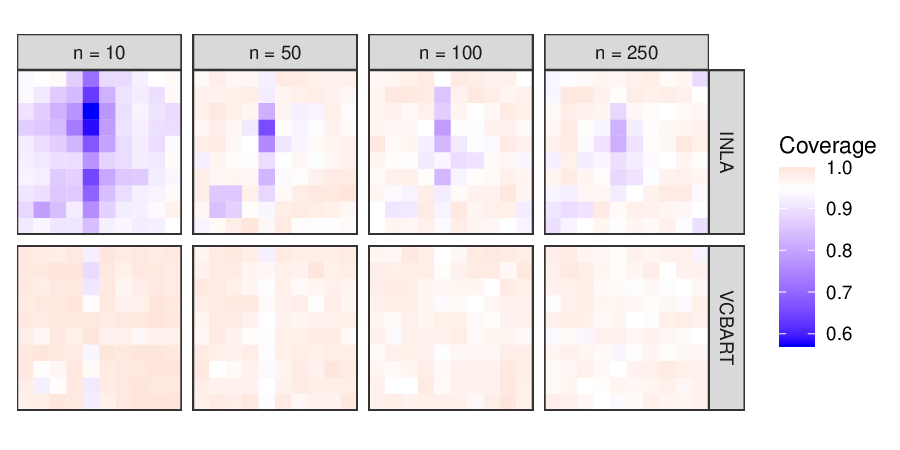}
    \caption{Average 95\% credible interval coverage across 200 simulations for $\Psi$ when sample size per cell is 10, 50, 100, and 250. Fixed settings: $\rho = 0, \sigma = 1$.}
    \label{fig:sim-stats-detail-psi}
\end{figure}

While the global statistics suggest the two models are performing at a somewhat similar level, there are differences in each model's ability to estimate the local mixture effects for each grid cell. The 95\% credible interval coverage for the local mixture effects is shown in Figure \ref{fig:sim-stats-detail-psi} for the high noise setting ($\sigma = 1$) with uncorrelated exposures ($\rho = 0$). The coverage for CAR is very poor for many of the cells in the lowest sample size ($n = 10$), but improves some as the sample size increases. On the other hand, VCBART generally has better coverage across all grid cells, with near 95\% coverage even in small sample sizes.

Spatial patterns of poor coverage for local mixture effects might be attributable to poor coverage for one or more of the constituent local exposure coefficients. We found that the CAR model struggles most with the spatial patterns used to generate $\beta_1$, $\beta_2$, and $\beta_6$ (see Figure S1.3 in the Supplementary Material for simulation average coverage for each spatially varying coefficient for the $n = 100$ setting, corresponding to the third column of Figure \ref{fig:sim-stats-detail-psi}). These surfaces contain some of the sharpest contrasts between neighboring cells, which presents difficulties for models which rely on spatial smoothing. VCBART also struggles to capture the two hot spots in the $\beta_6$ surface, but not to the same extent as the CAR model, and generally has as good or better coverage across the other parameters.

These results suggest that VCBART may be preferable to CAR in settings with high-noise or small local sample sizes. In general, we found that as $\rho$ increases, coverage and bias for $\Psi(z)$ improves or changes little, while coverage and bias for the spatially varying regression parameters worsens. The latter was particularly noticeable for the CAR model. We also noticed that coverage for the CAR model was substantially worse in the high noise variance setting, whereas the amount of noise had little effect on coverage for VCBART.

\section*{Application}
\label{sec:application}
In an application of the VCBART model, we analyzed 1,468,531 live singleton births to mothers residing in Georgia with an estimated conception date between January 1st, 2005 and December 31st, 2016. In this sample, the majority of mothers were white (58.1\%), and in terms of educational attainment about half (50.7\%) reported at least some college experience. Additional demographic information is provided in Table \ref{tab:maternal-dm}.

\begin{table}

\caption{\label{tab:maternal-dm}Maternal Demographic Characteristics}
\centering
\begin{tabular}[t]{>{\raggedright\arraybackslash}p{7cm}rr}
\toprule
Characteristic & N & \%\\
\midrule
\addlinespace[0.3em]
\multicolumn{3}{l}{\textbf{Race}}\\
\hspace{1em}White & 853,575 & 58.1\\
\hspace{1em}Black & 505,304 & 34.4\\
\hspace{1em}Asian or Pacific Islander & 58,706 & 4.0\\
\hspace{1em}American Indian or Alaskan Native & 2,460 & 0.2\\
\hspace{1em}Other & 48,486 & 3.3\\
\addlinespace[0.3em]
\multicolumn{3}{l}{\textbf{Ethnicity}}\\
\hspace{1em}Non-Hispanic & 1,252,268 & 85.3\\
\hspace{1em}Hispanic & 216,263 & 14.7\\
\addlinespace[0.3em]
\multicolumn{3}{l}{\textbf{Age}}\\
\hspace{1em}Less than 25 years & 519,590 & 35.4\\
\hspace{1em}25-31 years & 568,344 & 38.7\\
\hspace{1em}More than 31 years & 380,597 & 25.9\\
\addlinespace[0.3em]
\multicolumn{3}{l}{\textbf{Education}}\\
\hspace{1em}Less than 9th grade & 73,584 & 5.0\\
\hspace{1em}9th-11th grade & 204,226 & 13.9\\
\hspace{1em}12th grade & 446,449 & 30.4\\
\hspace{1em}Some college & 744,272 & 50.7\\
\bottomrule
\end{tabular}
\end{table}

We fit a VCBART model with the default 50-trees-per-ensemble setting. A spatially varying intercept, as well as spatially varying coefficients for quantized versions of PM$_{2.5}$, NO$_2$, SO$_2$, O$_3$, and CO were included. For this analysis, we chose to quantize each exposure into 10 quantile bins, i.e., deciles. Additional covariates modeled using fixed effects included estimated conception date, gestational age, tobacco use, and the parity, age, race, ethnicity, level of educational attainment, and marital status of the mother. We also adjusted for socioeconomic status using Census tract-level estimates of the percentage below the poverty level. The continuous covariates age, tract poverty level, and conception date were modeled used natural cubic splines with 5 degrees of freedom, while gestational age was modeled using indicator variables for the number of weeks.

\begin{figure}
    \centering
    \includegraphics[width=6.5in]{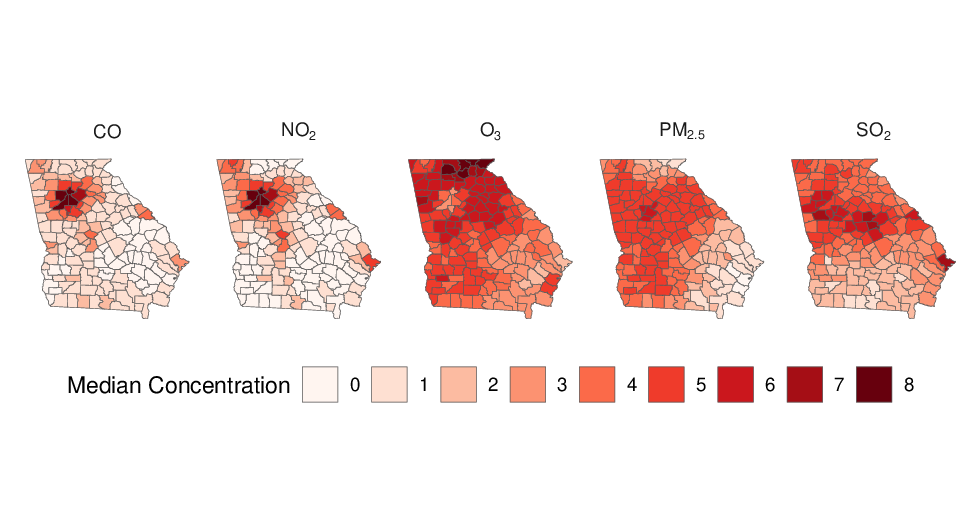}
    \caption{Median pregnancy-wide concentration for each pollutant after having been quantized into deciles.}
    \label{fig:app-quant-con-cty-map}
\end{figure}

As previously mentioned, one of the reasons a spatially varying coefficient model might be appropriate for this analysis is that pollutant concentrations may vary across space. We calculated the average (mean) pregnancy-wide concentration of each pollutant within each Georgia county. The distribution of these county-level averages are summarized in Table \ref{tab:pol-con-cty}. Most notably, mothers in counties at the 90th percentile of NO$_2$ and SO$_2$ were, on average, exposed to more than double the concentration of these pollutants compared to mothers in counties at the 10th percentile. Figure \ref{fig:app-quant-con-cty-map} displays the median pregnancy-wide pollutant concentrations for each county, after the exposures have been quantized into deciles. While pollutant concentration typically varies seasonally, some trends are clear, such as CO and NO$_2$ concentrations being highest in the Atlanta metropolitan area, and NO2 following the path of I-75.

\begin{table}

\caption{\label{tab:pol-con-cty}Percentiles of County-level Mean Pregnancy-wide Pollutant Exposures}
\centering
\begin{tabular}[t]{lrrrrr}
\toprule
\multicolumn{1}{c}{ } & \multicolumn{5}{c}{Percentile} \\
\cmidrule(l{3pt}r{3pt}){2-6}
Pollutant & 10th & 25th & 50th & 75th & 90th\\
\midrule
24-hr average PM$_{2.5}$ (µm/m$^3$) & 9.00 & 9.59 & 10.22 & 10.65 & 10.65\\
1-hr max NO$_2$ (ppb) & 4.40 & 5.12 & 6.60 & 9.53 & 9.53\\
1-hr max SO$_2$ (ppb) & 1.90 & 2.25 & 3.14 & 3.99 & 3.99\\
8-hr max O$_3$ (ppb) & 38.98 & 39.66 & 40.50 & 41.22 & 41.22\\
1-hr max CO (ppb) & 0.30 & 0.31 & 0.33 & 0.37 & 0.37\\
\bottomrule
\end{tabular}
\end{table}

\begin{figure}
    \centering
    \includegraphics[width=5in]{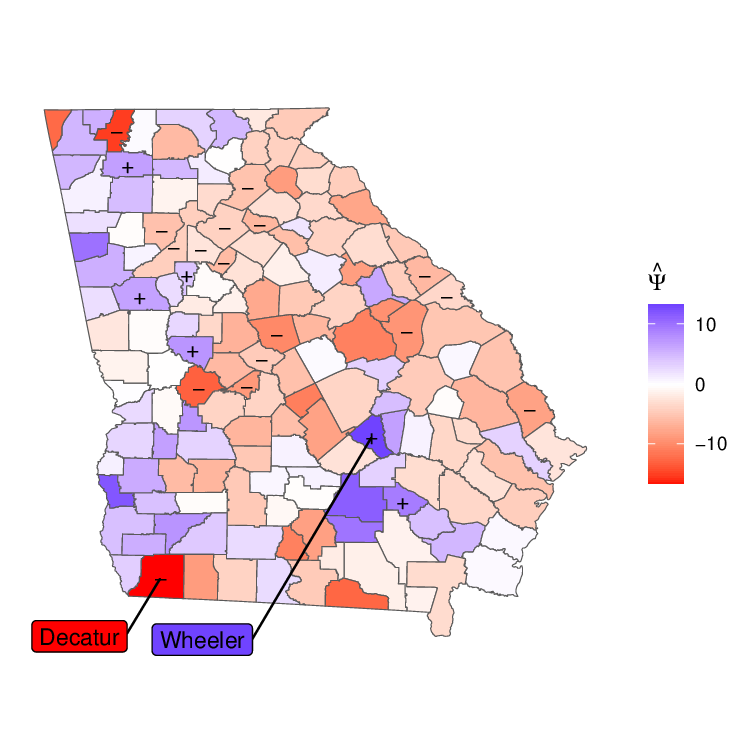}
    \caption{Local mixture effects estimated using VCBART. The mean of the posterior distribution of $\Psi(z_i)$ at each location $z_i$ is displayed. Counties marked with a ``--" have a 95\% credible interval entirely below zero, and counties marked with a ``+" have a 95\% credible interval entirely above zero.}
    \label{fig:app-vcbart-psi-map}
\end{figure}

Posterior means of the mixture effects using the VCBART model are plotted in Figure \ref{fig:app-vcbart-psi-map}. In general, we found stronger negative mixture effects in counties making up the central and eastern portion of the state. For a one decile increase in all five exposures, the county-specific estimates range from an expected reduction in birthweight of -16.65 grams (95\% CrI: -33.93, -0.40) in Decatur county to an increase of 13.28 grams (95\% CrI: 0.06, 27.19) in Wheeler county. Of the counties with 95\% credible intervals that exclude zero, 17/23 are in a negative direction. We have supplied a forest plot of these 23 county-level mixture effects in the Supplementary Material. As a comparison, we also estimated a common mixture effect using a linear model with only a spatially varying CAR intercept and no spatially varying coefficients. This common mixture effect was estimated to be a reduction of -1.81 grams (95\% CrI: -2.84, -0.70) per decile increase in all pollutants. This estimate is slightly attenuated compared to a weighted average of the local mixture effects from the VCBART model (reduction of 2.27 grams). VCBART outperformed the CAR intercept model, as well as a spatially varying coefficient CAR model akin to that which was fit in the simulation study, in terms of the Widely Applicable Information Criterion (see Table S2.1 in the Supplemental Material).\citep{watanabe_asymptotic_2010}

\begin{figure}
    \centering
    \includegraphics[width=5.5in]{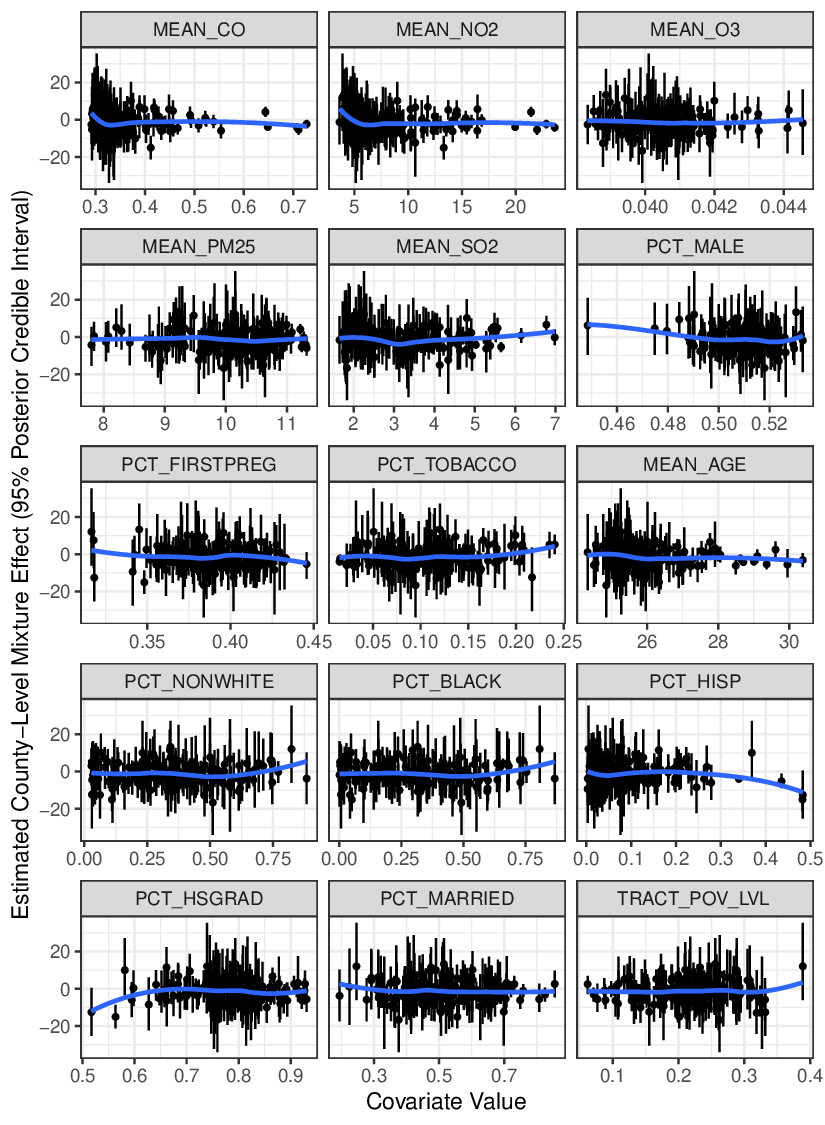}
    \caption{Local mixture effects estimated using VCBART compared to selected county-level average exposures and confounders with a smooth loess curve overlaid.}
    \label{fig:app-vcbart-psi-scatter}
\end{figure}

In Figure \ref{fig:app-vcbart-psi-scatter}, we plot the estimated local mixture effects against county-level summaries of the exposures and confounders included in the VCBART model. For the most part, there is no discernible pattern in the estimated mixture effects when compared to the confounders. For CO and NO2$_2$, the local mixture effects tend to shift in the negative direction as the level of exposure increases at the lower end of the observed concentrations.

\section*{Discussion}
\label{sec:discussion}

We describe varying coefficient models as a useful extension to the popular QGCOMP method to account for when heterogeneous exposure-response relationships are present in the data. We have shown through simulation and an analysis of birth records in Georgia how one might estimate spatially varying parameters in such a model via a Bayesian approach which uses CAR or BART priors. In our analysis of birth records, we found that for many Georgia counties there exists an association between elevated concentrations of a mixture of PM$_{2.5}$, NO$_2$, SO$_2$, O$_3$, and CO and reduced birthweight.

VCBART is not the only option for fitting the model we have described, as there is a rich literature of spatially varying coefficient models. \citet{casetti_generating_1972} originally described an expansion method for generating improved models by taking the parameters from an initial model and making them a function of variables. Later the geographically weighted regression (GWR)\citep{fotheringham_geographically_2002} and aforementioned SVC\citep{gelfand_spatial_2003} models were proposed. GWR is a frequentist approach that involves estimating a separate weighted least squares regressions at each location, where the weights are determined by proximity between locations as measured by some kernel function. The SVC model is a Bayesian approach which places Gaussian process priors on the individual regression coefficients, where again a kernel function is used to estimate the distance between observations. Comparisons of the two approaches have found similar performance in many settings, but note that GWR may occasionally struggle in the presence of correlated covariates.\citep{wheeler_assessment_2007, wheeler_comparing_2009, finley_comparing_2011} The SVC model provides a richer framework for making predictions on new spatial locations, which might make it a good option for estimating smooth mixture effects over a region from point-referenced data.

A limitation of the analysis is the measurement of exposures. Not only is there potential for error in the exposure measurements themselves, but the mechanism by which we assign pregnancy-averaged pollutant concentrations to each mother is imperfect. The residential address on file may not be reflective of where the mother spent most her time during the pregnancy, and even when it is, the amount of exposure two individuals from the same neighborhood experience could be very different due to unmeasured factors such as occupation or personal lifestyle behaviors.

Generally, ambient air pollution contributes little explanatory power for birthweight. The $R^2$ value for the VCBART model in the application is 39.90\%. It is possible that all or some subset of these pollutants may be more informative if averaged during a critical window of the pregnancy instead of the entire duration. Previous studies have observed different associations between air pollution and birthweight in Atlanta, Georgia for specific months or trimesters of pregnancy.\citep{darrow_ambient_2011, strickland_associations_2019} Various data driven methods have been developed for identifying windows of pregnancy particularly susceptible to air pollution, including some based on BART. \citep{wilson_potential_2017, chang_assessment_2015, mork_estimating_2023, mork_heterogeneous_2023} On a similar note, spatially varying distributed lag models, such as the one proposed in \citet{warren_spatially_2020}, might be another QGCOMP extension worth exploring.

In terms of the methods proposed, VCBART is more computationally burdensome than the CAR model, particularly when an efficient implementation like INLA is used. This is due to the overhead required for managing tree structures. However, we have found that this tree-based approach is advantageous over the CAR model to estimate local mixture effects, particularly in high-noise settings such as our birthweight analysis. In this work, we focused on a spatially heterogeneous approach to quantile g-computation that made use of VCBART's graph-structured branching process. However, in practice one could also use the traditional BART branching process to model heterogeneity in the exposure coefficients as a function of demographic or clinical covariates as in \citet{englert_estimating_2025}. For instance, \citet{darrow_ambient_2011} reported higher estimates of the associations between various pollutants and birthweight for Hispanic and non-Hispanic black infants compared to non-Hispanic white infants. The current implementation of the \texttt{VCBART} R package requires the BART ensembles to all use the same set of covariates, and a future extension would also be to select different covariates for each exposure.


\bibliographystyle{unsrtnat}
\bibliography{references}

\end{document}


\maketitle

\tableofcontents

\section{Additional Simulation Study Materials}
\subsection{Details for Simulating Spatially Varying Coefficient Surfaces}

For the simulation study in the manuscript, we generate a spatially varying intercept and six spatially varying regression coefficients across a 10 x 10 grid using the following functions:
\begin{gather}
    \beta_0 (x, y) = 100 \phi \left( (x, y)^T \mid (5.5, 5.5)^T, 20 \mathbf{I}_2 \right) \\
    \beta_1(x, y) = 0.50 \times \mathbb{I}_{\left\{x > 5\right\}} \\
    \beta_2(x, y) = -0.25 \times \mathbb{I}_{\left\{y > 5\right\}} \\
    \beta_3(x, y) = -\exp{\left(-|x - 5.5|\right)} \\
    \beta_4(x, y) = \exp{\left(-|y - 5.5|\right)} \\
    \beta_5(x, y) = 50 \times \phi \left((x,y)^T \mid (1,1)^T, 10\mathbf{I}_2 \right) \\
    \beta_6(x, y) = 4 \times \left[ \phi \left((x,y)^T \mid (7.5,7.5)^T, \mathbf{I}_2 \right) + \phi \left((x,y)^T \mid (2.5,2.5)^T, \mathbf{I}_2 \right) \right]
\end{gather}
where $x$ and $y$ correspond to the integer dimensions of the cells in the grid, $\phi$ is the probability density function of a bivariate normal distribution, $\mathbb{I}$ is an indicator function, and $\mathbf{I}$ is a diagonal identity matrix.


\begin{figure}[h]
    \centering
    \includegraphics[width=\linewidth]{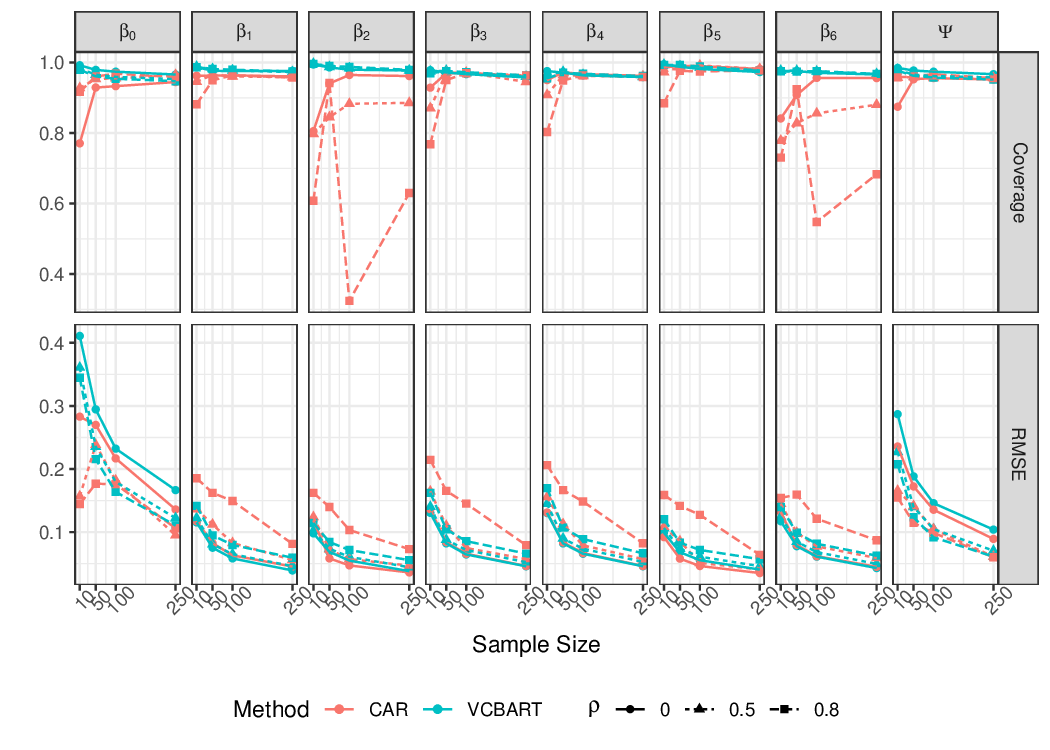}
    \caption{Global average 95\% credible interval coverage and root mean squared error (RMSE) for the mixture effect $\Psi$ and individual exposure coefficients (high noise setting). Global values are calculated using the following approach: $\text{Coverage} = \frac{1}{B} \sum_{b=1}^{B} \left(\frac{1}{100}\sum_{z} \mathbb{I}_{\left\{\Psi(z) \in [\hat{\Psi}(z)_{0.025}, \hat{\Psi}(z)_{0.975}] \right\}} \right)$, and $\text{RMSE} = \frac{1}{B} \sum_{b=1}^{B} \left(\frac{1}{100}\sum_{z} (\Psi(z) - \hat{\Psi}(z))^2 \right)$. Fixed settings: $\sigma = 1$.}
    \label{fig:sim-summary-beta-1-all-6}
    \addcontentsline{toc}{subsection}{Figure \thefigure: Global Coverage and RMSE for All Parameters (high noise setting)}
\end{figure}

\begin{figure}[h]
    \centering
    \includegraphics[width=\linewidth]{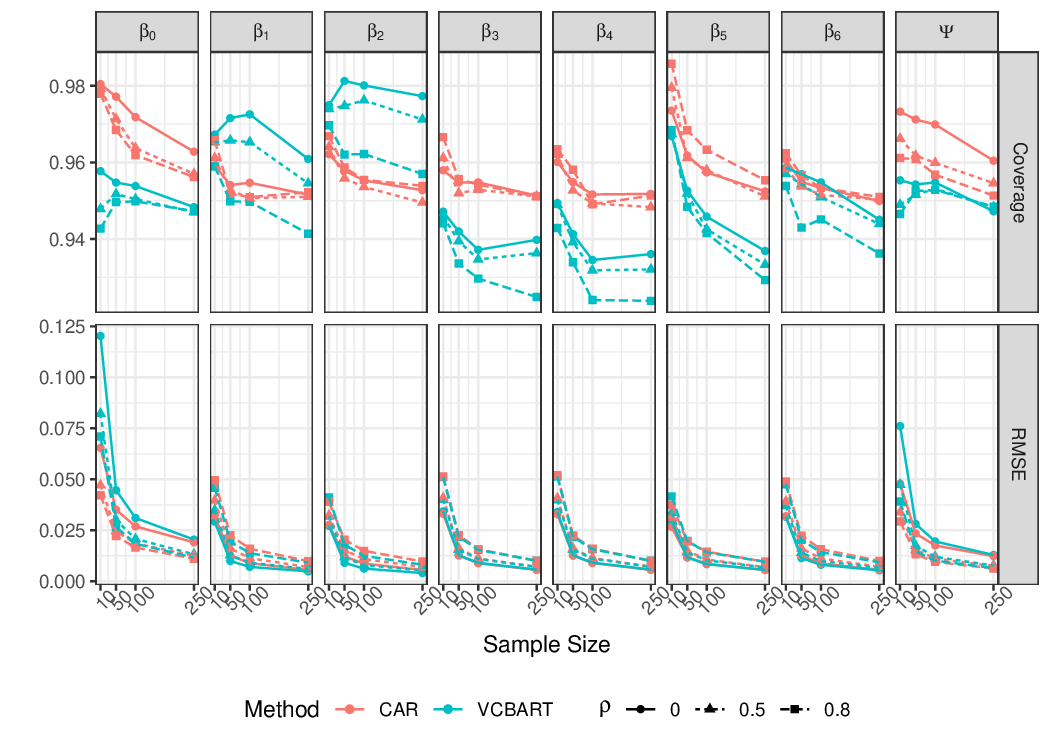}
    \caption{Global average 95\% credible interval coverage and root mean squared error (RMSE) for the mixture effect $\Psi$ and individual exposure coefficients (low noise setting). Global values are calculated using the following approach: $\text{Coverage} = \frac{1}{B} \sum_{b=1}^{B} \left(\frac{1}{100}\sum_{z} \mathbb{I}_{\left\{\Psi(z) \in [\hat{\Psi}(z)_{0.025}, \hat{\Psi}(z)_{0.975}] \right\}} \right)$, and $\text{RMSE} = \frac{1}{B} \sum_{b=1}^{B} \left(\frac{1}{100}\sum_{z} (\Psi(z) - \hat{\Psi}(z))^2 \right)$. Fixed settings: $\sigma = 1$.}
    \label{fig:sim-summary-beta-0.1-all-6}
    \addcontentsline{toc}{subsection}{Figure \thefigure: Global Coverage and RMSE for All Parameters (low noise setting)}
\end{figure}

\begin{figure}[h]
    \centering
    \includegraphics[width=6in]{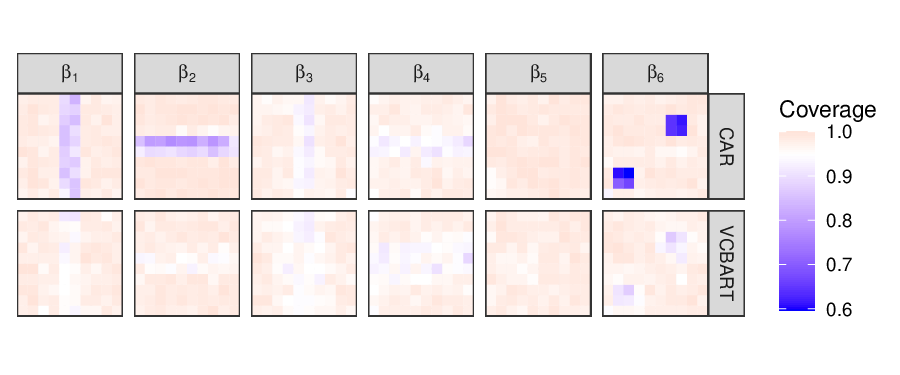}
    \caption{Average 95\% credible interval coverage across 200 simulations for all spatially varying regression coefficients. Fixed settings: $n = 100, \rho = 0, \sigma = 1$.}
    \label{fig:sim-stats-detail-beta}
    \addcontentsline{toc}{subsection}{Figure \texorpdfstring{\thefigure: Local Coverage for All Coefficients ($n = 100, \rho = 0, \sigma = 1$)}{XX}}
\end{figure}

\clearpage

\section{Additional Application Materials}

\begin{figure}[h]
    \centering
    \includegraphics[width=5in]{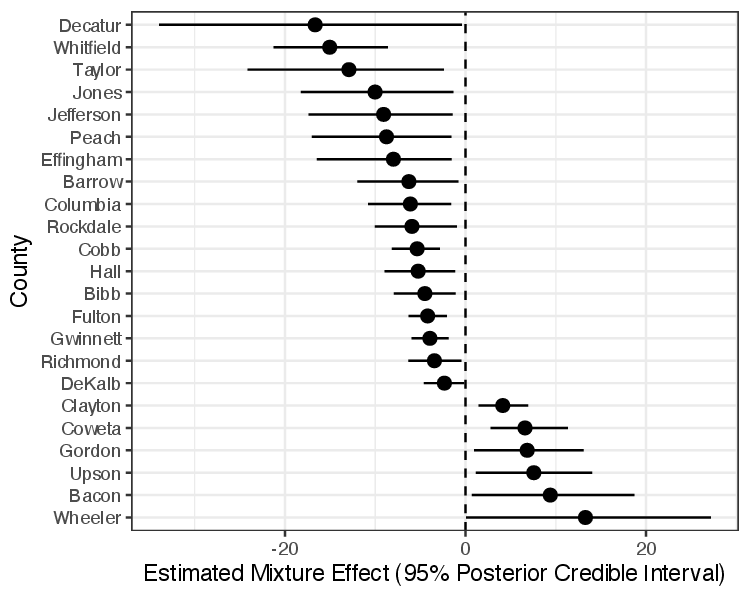}
    \caption{Local mixture effects estimated using VCBART. Only the 23 counties with 95\% posterior credible intervals excluding zero are shown.}
    \label{fig:app-vcbart-psi-forest}
    \addcontentsline{toc}{subsection}{Figure \thefigure: Forest Plot of Local Mixture Effect Estimates}
\end{figure}

\begin{table}[h]

\caption{\label{tab:app-relative-waic}WAIC for Candidate Models}
\centering
\begin{tabular}[t]{lr}
\toprule
Model & WAIC\\
\midrule
VCBART & 21,937,450\\
OLS & 21,939,763\\
VC CAR & 21,943,136\\
RI CAR & 21,943,089\\
\bottomrule
\multicolumn{2}{l}{\rule{0pt}{1em}WAIC: Widely applicable information criterion.}\\
\multicolumn{2}{l}{\rule{0pt}{1em}OLS: Non-spatial ordinary least-squares regression.}\\
\multicolumn{2}{l}{\rule{0pt}{1em}VC CAR: Spatially varying coefficient CAR model.}\\
\multicolumn{2}{l}{\rule{0pt}{1em}RI CAR: Spatial random intercept CAR model.}\\
\end{tabular}
\addcontentsline{toc}{subsection}{Table \thetable: WAIC for Candidate Models}
\end{table}